\documentclass[twocolumn,aps,showpacs,prl]{revtex4}

\usepackage{graphicx}
\usepackage{amssymb}
\usepackage{color}

\begin{document}

\newcommand{\be}{\begin{equation}}
\newcommand{\ee}{\end{equation}}
\newcommand{\bea}{\begin{eqnarray}}
\newcommand{\eea}{\end{eqnarray}}
\newcommand{\Xing}{{\color{red}Xing-comment: }}

\title{Topological Defects in Spherical Nematics}
\author{Homin Shin, Mark J. Bowick, and Xiangjun Xing}
\affiliation{Department of Physics, Syracuse University, Syracuse NY
13244-1130, USA}

\begin{abstract}
We study the organization of topological defects in a system of
nematogens confined to the two-dimensional sphere ($S^2$). We first
perform Monte Carlo simulations of a fluid system of hard rods
(spherocylinders) living in the tangent plane of $S^2$. The sphere
is adiabatically compressed until we reach a jammed nematic state
with maximum packing density. The nematic state exhibits four $+1/2$
disclinations arrayed on a great circle rather than at the vertices
of a regular tetrahedron. This arises from the high elastic
anisotropy of the system in which splay ($K_1$) is far softer than
bending ($K_3$). We also introduce and study a lattice nematic model
on $S^2$ with tunable elastic constants and map out the preferred
defect locations as a function of elastic anisotropy. We establish
the existence of a one-parameter family of degenerate ground states
in the extreme splay-dominated limit $K_3/K_1 \rightarrow \infty$.
Thus the global defect geometry is controllable by tuning the
relative splay to bend modulus.

\end{abstract}
\maketitle

Ordered arrays of microscopic structures on curved interfaces
provide a promising route to fabricating nanoscale or mesoscale
building blocks (mesoatoms) that may in turn form molecules and bulk
materials. One class of mesoatoms is provided by particles
self-assembling on spherical droplets in liquid-liquid emulsions.
The particles may be isotropic or shaped. Ordered structures on
spherical interfaces always possess topological defects
~\cite{Nelsonb}. The location and detailed arrangement of such
defects are both important since defects are distinctive regions
which may be functionalized to create directional bonds akin to
atomic bonds. This has been nicely illustrated recently by the
synthesis of divalent gold nanoparticles coated with self-assembled
stripes of phase-separated ligands whose two polar defects can be
functionalized. The divalent nanoparticles subsequently link
spontaneously to form chains~\cite{DeVries}.

Spherical nematics have a local 2-fold inversion symmetry and
elementary disclinations of both half-integer and integer strength.
Since the total disclination strength on the 2-sphere is two
~\cite{LP,Nelson}, nematic ground states may possess four $+1/2$
defects, two $+1$ defects or two $+1/2$s and one $+1$.
Functionalization of the defects in the first case could lead to
tetravalent mesoatoms with sp$^3$-like directional bonding
~\cite{Nelson}. The structure and arrangement of defects in such a
thin nematic shell with variable thickness has recently been studied
both experimentally~\cite{FN:2007} and theoretically~\cite{VN:2006}.
The analyses so far have, however, been limited to the one Frank
constant approximation, in which the bending stiffness $K_3$ and the
splay constant $K_1$ are equal~\cite{Zannoni}. Although fluctuations
always drive these elastic moduli to the same value in the long
wavelength limit, we are necessarily working at finite volume on the
compact 2-sphere. It is essential, therefore, to explore the effect
of differing bend and splay moduli on the structure of defects in
the ground state.

To this end we have performed Monte Carlo simulations of hard rod
fluids confined to the tangent plane of the 2-sphere. Adiabatically
shrinking the sphere increases the packing density and leads to a
jammed state with nematic order and four $+1/2$ disclinations. Since
like-sign defects repel, one might expect them to be maximally
separated at the vertices of a regular tetrahedron~\cite{LP,Nelson}.
We find instead that the four defects lie approximately on a great
circle. This can be understood as arising from the high bending
stiffness $K_3$ compared to the splay stiffness constant $K_1$. We
also analyze a coarse-grained model of a spherical nematic with
tunable Frank constants and map out the global pattern of defects as
a function of the anisotropy $K_3/K_1$. In the limit $K_3/K_1
\rightarrow \infty$, we show that the system exhibits a
one-dimensional continuum of degenerate ground states in which the
four defects form a rectangle of arbitrary aspect ratio.

Our Monte Carlo simulation is performed in the isobaric-isothermal
(constant-NPT) ensemble with rod-like particles interacting via hard
core repulsion~\cite{Allen}. We use spherocylinders with length $L$
and diameter $D$ (aspect ratio $L/D$). Isotropic initial states of
$N$ rods are prepared by centering each randomly oriented rod on a
node of a spherical mesh of a 2-sphere of initial radius $R_i$. We
perform $\sim 10^7$ Monte Carlo cycles, each of which consists of a
translational and orientational trial move for all $N$ rods together
with a \emph{volume-compression} move. The compression rate must be
low enough to avoid premature jamming at low density. The simulation
stops when the system is jammed and thus cannot be further
compressed.

\begin{figure}[h!]
\centering
\includegraphics[angle=-90, scale=0.48]{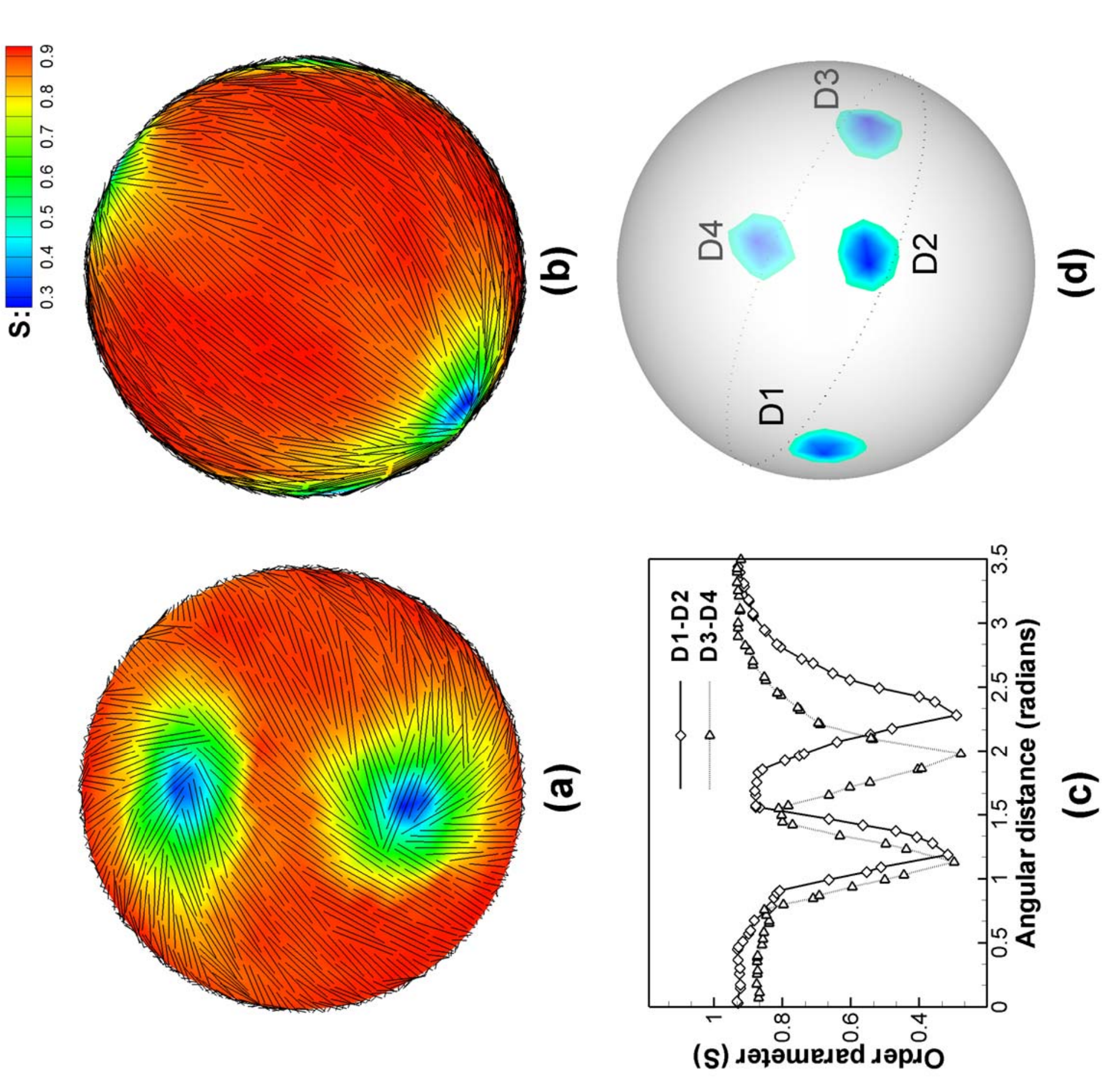}
\caption{(Color online) The ground-state configuration for spherical
nematic ordering of 1082 rods with $L/D=15$ is shown as the central
lines of rods for clarity, together with maps of the local nematic
order parameter. The two views are: (a) near defects and (b) nematic
bulk. The radius of the compressed sphere is $R_f/D=37.15$. The
measurement of the order parameter along the great circle connecting
defects is plotted as a function of angular distance in radians in
(c). The exact locations of the four defects lying near one great
circle are indicated in (d).} \label{order}
\end{figure}

We simulated a system of $N=1082$ rods with aspect ratio $L/D=15$.
The final radius reached was $R_f/D=37.15$ starting from
$R_i/D=100$. The final states are found to be nematic with four
$+1/2$ disclinations.  This contrasts with the corresponding
infinite system in 2D flat space where the densest packing state is
crystalline~\cite{BatesFrenkel}. At these system sizes, therefore,
the spatial curvature of the surface frustrates crystalline
ordering. Crystalline order is nevertheless expected to set in for
larger system sizes, where the radius of curvature becomes much
larger than the rod length. In simulations with short rods ($L/D <
7$) we find, in contrast, smectic-like domain structures at maximum
density.

To characterize the nematic order we measure the local order
parameter tensor $Q_{\alpha\beta}$, a symmetric, traceless second-
rank tensor, by locally averaging the orientation of $m(V)$ rods
$\hat{n}(a)$ within a small domain $V$ (typically with an area of
order $L^2$): \be \label{Q-def}
Q_{\alpha\beta}=\frac{1}{m(V)}\sum_{a \in V}
\left(\hat{n}(a)_{\alpha}\hat{n}(a)_{\beta}
-\frac{1}{2}\delta_{\alpha\beta}\right). \ee The nematic order
parameter $S$ is defined to be twice the positive eigenvalue of
$Q_{\alpha\beta}$. Maps of $S$ are shown in Figs.~\ref{order}(a) and
(b). The local order parameter along the great circle connecting the
defects is also plotted versus angular distance in
Fig.~\ref{order}(c). The maximum value of $S$ is about $0.93$, in
contrast with flat space where $S$ reaches $1$~\cite{BatesFrenkel}.
The reduced maximum on the sphere results from the frustration due
to the nonzero curvature enclosed in the region $V$. The defect
cores are identified as the loci of minima of the order parameter.
Four $+1/2$ disclinations are clearly observed to lie on a single
great circle, to within $\pm 0.15$rad, as illustrated in
Fig.~\ref{order}(d).

The planarity of the defect array can be explained by the strong
elastic anisotropy. It is known that, for hard rod systems, the bend
constant $K_3$ diverges as the rod density increases while the splay
constant $K_1$ is almost independent of
density~\cite{LeeMeyer,LeeMeyer2} . This leads to $K_3 \gg K_1$ at
high packing density. To check this we determine the Frank constants
by fitting the local director field surrounding a defect to a
numerical minimization of the Frank free energy~\cite{Hudson}. We
find that $K_3/K_1 \gtrsim 20$.

\begin{figure}[b]
\centering
\includegraphics[angle=0, scale=0.35]{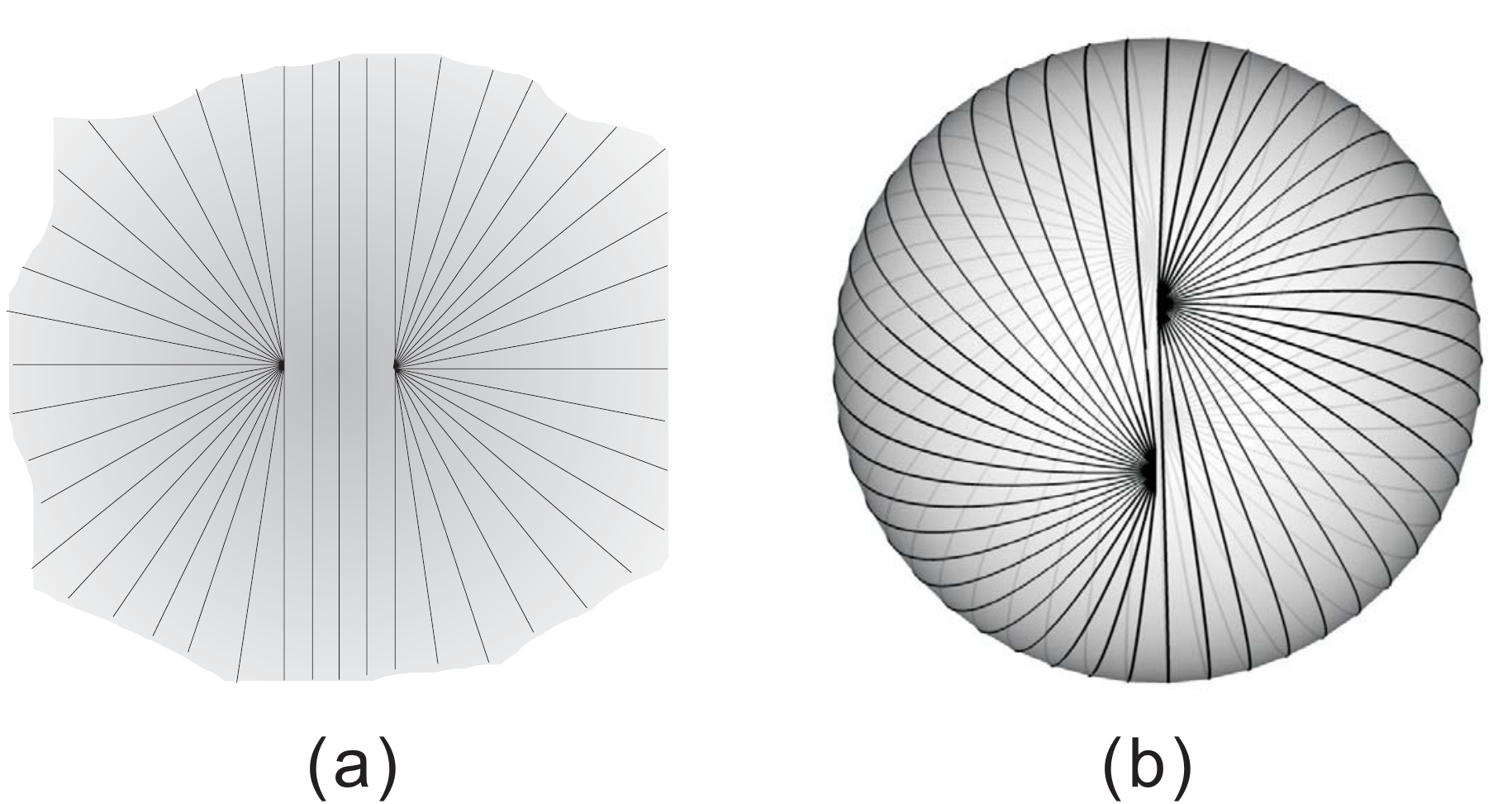}
\caption{(Color online) (a) In the limit of infinite $K_3/K_1$, the
separation of a $+1/2$ disclination pair in the plane costs no
energy. (b) On the 2-sphere, an infinite number of states with four
$+1/2$ disclinations on a great circle can be generated from a state
with two $+1$ disclinations by cut-and-rotate surgery.}
\label{+1dis}
\end{figure}

Consider first planar nematics in the strong anisotropy limit. It is
known that $+1$ disclinations admit both pure bending texture or
pure splay texture~\cite{Deem,note1}. When the bending and splay
constants are different one of these two textures will correspond to
a minimum of the energy and the other a maximum. The energy of a
$+1$ disclination is therefore given by $F^{(+1)} \sim \pi
\min(K_1,K_3) \ln (R/a_0)$, where $R$ is the system size and $a_0$
is the defect core size. The energy of a $+1/2$ defect, by contrast,
depends on both $K_1$ and $K_3$. In the one Frank constant limit
$K_1 = K_3 = K$, the energy of a $+1/2$ disclination is one-quarter
that of a $+1$ disclination: $F^{(+\frac{1}{2})} \sim \frac{\pi
K}{4} \ln (R/a_0)$. A $+1$ disclination will therefore unbind to a
pair of $+1/2$ disclinations which may then separate. In the
splay-dominated case $K_1 \ll K_3$, in contrast, the energy of
$+1/2$ disclination is half that of a $+1$ disclination:
$F^{(+\frac{1}{2})} \sim \frac{\pi K_1}{2} \ln (R/a_0)$. Splitting a
$+1$ defect therefore gains no energy and the total energy of two
$+1/2$ disclinations is independent of their separation. All bending
modes are frozen for a 2D nematic in flat space with $K_3 = \infty$.
It can be shown in this limit that the integral curve of the
director field is a geodesic (straight line). This implies that the
2D nematic director texture is completely determined by its
configuration on a one-dimensional curve: there is no bulk degree of
freedom that one can vary. In Fig.~\ref{+1dis}(a) we illustrate the
formation and separation of a pair of $+1/2$ pure-splay defects by
{\em global} surgery on a $+1$ pure-splay defect. The $+1$ defect is
pulled apart horizontally and the intermediate region filled with
uniform nematic texture. Clearly it is also possible to pull apart
the defect pair vertically without free energy cost.  By contrast,
any infinitesimal {\em local} deformation of the state in
Fig.~\ref{+1dis}(a) generates bending deformations and is therefore
forbidden energetically.

We now extend this argument to nematic order on the 2-sphere with
infinite bending stiffness. The constraint of no bending again
translates into the requirement that the integral curves of the
director follow geodesic lines (great circles). Consider an initial
state for which the director field follows lines of longitude
everywhere. This state, with one $+1$ disclination at each pole, is
clearly bending-free and is therefore a ground state in the limit
$K_3 \rightarrow \infty$.  Any local deformation of this state
introduces bend and is therefore forbidden. There are, however,
global manipulations that cost zero energy. As illustrated in
Fig.~\ref{+1dis}(b), we can cut the sphere into two hemispheres
along a great circle that contains both $+1$ disclinations, and
rotate one hemisphere by an arbitrary angle $\alpha$. This surgery
divides each $+1$ disclination into two $+1/2$ disclinations and the
resultant four $+1/2$ disclinations all lie on the great circle
along which we cut. Note that the director field is everywhere
smooth up to first order derivatives on the cut after the
surgery~\cite{note2}, except at the disclination cores. The elastic
free energy of the post-surgery state is therefore independent of
the rotation angle $\alpha$. We have thus identified a continuous
manifold of degenerate low energy states, parameterized by the angle
$\alpha$, for spherical nematics in the pure-splay limit $K_3
\rightarrow \infty$. The energy difference between the four $+1/2$
disclination state and the original state with two $+1$
disclinations can only come from the defect core energies and these
do not grow with radius $R$. It is thus negligible for large system
sizes.

Starting from a pure splay state and rotating the director at every
point by $\pi/2$ in the tangent plane, we obtain a pure bend state.
It can also be shown that the splay and bending constants $K_1$ and
$K_3$ interchange under this rotation. A system with $K_3/K_1 \ll 1$
should, therefore, also exhibit a similar one-parameter family of
degenerate ground states with a co-planar configuration of defects.
This is verified below. We note that the same cut-and-rotate surgery
has been applied to spherical smectics (for which $K_3/K_1$ is also
$ \gg 1 $) by Chantawansri et al;~\cite{Kramer-bc-sphere} and Blanc
and Kleman~\cite{Kleman}.


\begin{figure}[t]
\centering
\includegraphics[angle=-90, scale=0.38]{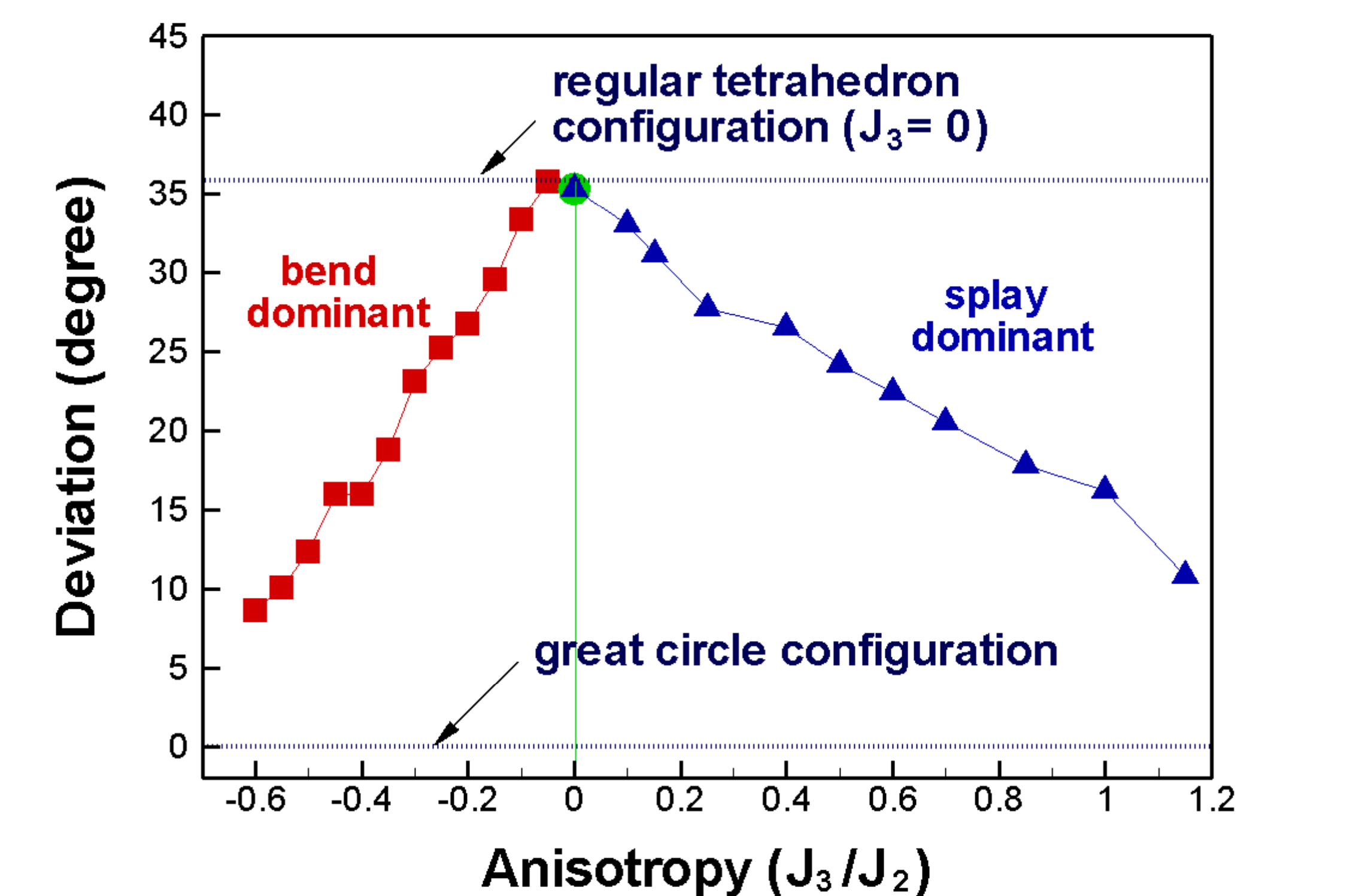}
\caption{(Color online) The effect of the anisotropy $J_3/J_2$ on the configuration
of defects. As the anisotropy is turned on the defects shift from a
tetrahedral geometry to a coplanar great circle alignment. The
angular deviation of the defect positions from the great circle are
measured as a function of the anisotropy.} \label{location}
\end{figure}

To understand better how the global configuration of defects depends
on the elastic anisotropy $K_3/K_1$, we next analyze a lattice
nematic model on $S^2$ in which the two Frank constants can be
continuously varied. We first construct a triangular lattice on the
2-sphere with $N=762$ vertices. As a result of the Gauss-Bonnet
theorem, twelve of these vertices have coordination number five and
form a regular icosahedron. It turns out, however, that these
crystalline disclinations do not significantly affect the location
of the nematic defects\cite{note3}. On every vertex $a$ of this
lattice we define a nematic director $\hat{n}_a$, which is
constrained to the local tangent plane.  The Hamiltonian is given by
\bea
\label{discret}
&& H = J_2 \sum_{\langle a,b\rangle} {\rm Tr}
    (N_a - N_b)  (N_a - N_b) \\
&+ & J_3 \sum_{ \langle a,b\neq c \rangle}
(\hat{e}_{ab} \cdot N_{a} \cdot  \hat{e}_{ac})
 {\rm Tr}  (N_a - N_b) \cdot (N_a - N_c) ,
\nonumber \eea where $N_a$ is the dyadic tensor $\hat{n}_a\hat{n}_a$, while
$\hat{e}_{ab}$ is the unit vector pointing from vertex $a$ to vertex
$b$. In Eq.~(\ref{discret}) $ \langle a,b \rangle$ means summation
over all pairs of nearest-neighbor sites $b$ and $c$ that are both
nearest neighbors of site $a$.

The lattice model Eq.~(\ref{discret}) is manifestly invariant under
spatial inversion of each director $\hat{n}_a$.  The two body term
with coefficient $J_2$ is precisely the 2D Maier-Saupe lattice model
of nematics, which is known to reduce, at large scales, to the Frank
free energy with equal elastic constants $K_1 = K_3$. The three body
term with coefficient $J_3$ couples the directors explicitly to the
underlying lattice and renders $K_1$ and $K_3$ different. It can be
shown~\cite{unpublished} that at large scales the lattice model
Eq.~(\ref{discret}) reduces to the following Frank free energy for
spherical nematics: \bea \label{F-frank} F= \int \sqrt{g}\, d^2{\bf
r} \, \left[
    \frac{K_1}{2} (\vec{D} \cdot  \hat{n} )^2
+ \frac{K_3}{2} ( \vec{D} \cdot \hat{t })^2 \right], \eea where
$\vec{D}$ is the covariant derivative and $\hat{t}$ is the unit
tangent vector perpendicular to the director $\hat{n}$. We note that
in going from Eq.~(\ref{discret}) to Eq.~(\ref{F-frank}), however,
terms of higher order (in $D \hat{n}$) have been dropped. These
terms, though irrelevant in the renormalization group sense, may be
quantitatively important for small system sizes. The parameter $J_3$
in Eq.~(\ref{discret}) turns out to be proportional to $(K_3 -
K_1)/(K_3 + 2 K_1)$. The coefficient of proportionality, however,
depends on microscopic details~\cite{unpublished}.

Starting from a random state, we determine the ground state of the
model Eq.~(\ref{discret}) using simulated annealing. We use the same
method to identify the nematic defects as we used for the hard rods
simulation. We first simulate the one Frank constant case for which
$J_3 = 0$ and find that the four defects form a regular tetrahedron
as expected~\cite{Nelson,LP}. A non-zero three-body coupling $J_3$
leads to many metastable states. To avoid them, we start from the
minimum energy configuration for the one Frank constant case and
turn on the parameter $J_3$ at sufficiently low temperature. We then
cool down again to reach the ground state. We find that the nematic
texture becomes increasingly splay (bend) dominated as $J_3$
increases (decreases) from zero. We also measure the angular
deviations of the defect locations from the circumscribed great
circle, determined by a least squares fit. As shown in
Fig.~\ref{location}, the defects gradually shift toward a great
circle as $|J_3|$ increases. When $J_3/J_2 \gtrsim  1.2 \,\,{\rm or} \,\, \lesssim
-0.6 $, the director field develops a local instability, from which
we infer that the splay or bend constant (at short scales) becomes
negative.  Nevertheless, the defect array is not strictly planar at
$J_3/J_2 \approx 1.2 \,\,{\rm or} \,\, -0.6 $. We suspect that this is due
to the higher order terms that have been dropped in going from the
discrete model Eq.~(\ref{discret}) to the continuous model
Eq.~(\ref{F-frank}), which in principle can renormalize the two
Frank constants and make them scale dependent. Such effects,
however, are expected to diminish at large system sizes.

\begin{figure}[t]
\centering
\includegraphics[angle=-90, scale=0.42]{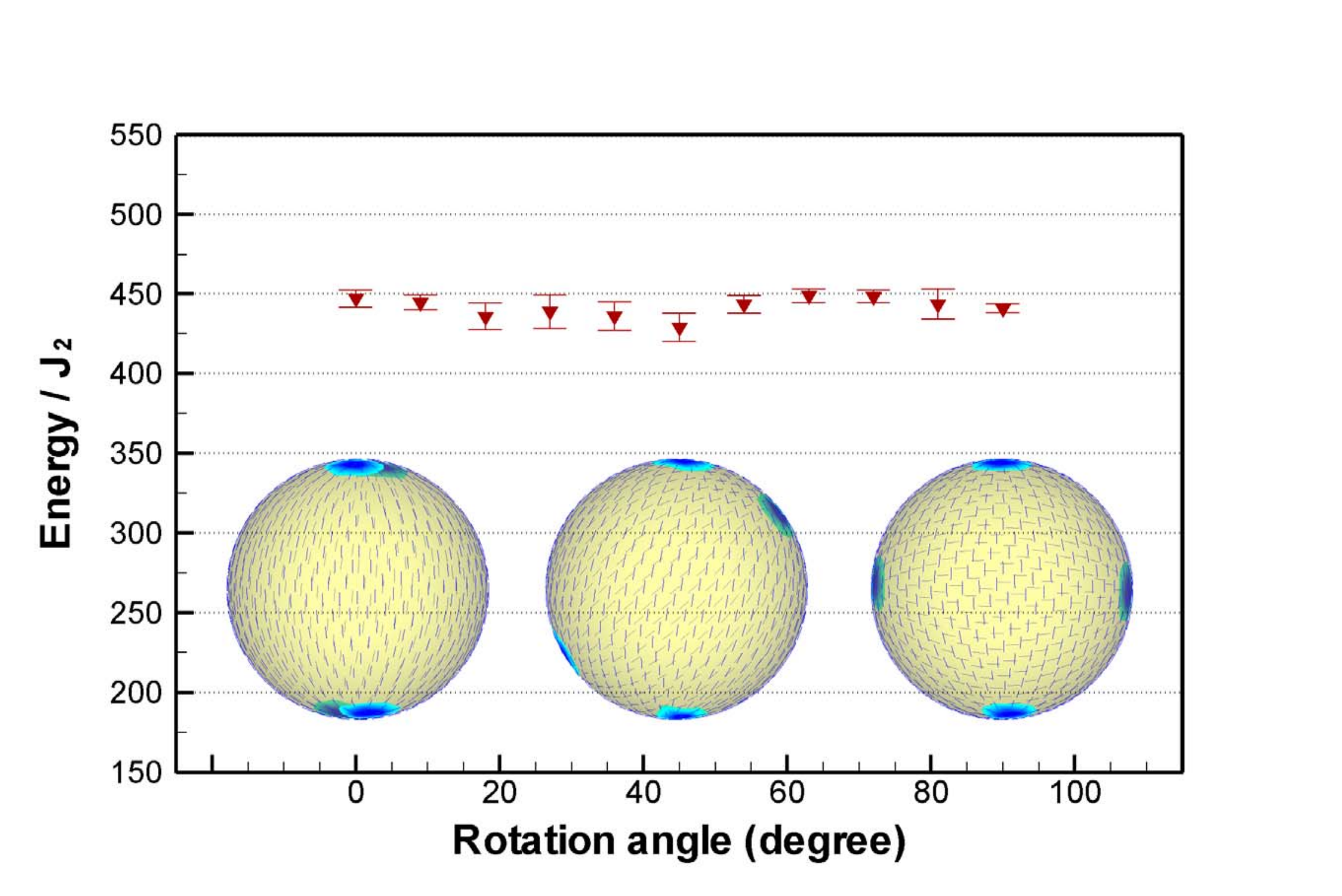}
\caption{(Color online)  At $J_3/J_2 = 1.2$, the energy, as a
function of the rotation angle $\alpha$, is flat within $\pm 2 \%$.}
\label{rotation}
\end{figure}

The lattice nematic model can be used to display explicitly the 1D
manifold of degenerate ground states. For $J_3/J_2 = 1.2$, we start
with a low energy state obtained by simulation of
Eq.~(\ref{discret}) in which the four defects lie approximately on
the equator with an angular separation of $90^{\circ}$. We then
force two defects to move within the great circle by rotating all
the directors in the lower hemisphere. Starting from this new state
we then minimize Eq.~(\ref{discret}) again at zero temperature. As
shown in Fig.~\ref{rotation}, the energy of all low energy states
thus obtained, for different angular separation $\alpha$, agree to
within $\pm 2 \%$. This small difference is likely due to finite
size effects.

In this letter we have studied the global configuration of defects
in the ground state of a spherical nematic using two different
models. We show that the tetrahedral configuration crosses over to a
{\em great-circle} configuration as we increase the anisotropy of
the elastic constants. Our work therefore explicitly demonstrates
that defect positions can be controlled by varying the elastic
anisotropy. This result should be relevant to designing and
fabricating mesoscopic molecules and bulk materials by attaching
ligands to functionalized defect sites. The work of MJB and HS was
supported by the NSF through Grants DMR-0219292 and DMR-0305407. The
work of XX was supported by the American Chemical Society through
Grant PRF 44689-G7. MJB thanks Robin Selinger for discussions.


\begin{thebibliography}{10}
\bibitem{note2}{ Other cut-and-rotate manipulations
either do not change the initial state or generate nematic director
discontinuities, which are energetically costly. }
\bibitem{note3} {We have also performed simulations on a random
mesh on the 2-sphere and obtain similar results. }
\end{thebibliography}
\end{document}